\documentclass[aps,prl,12pt,amsmath,amssymb,amsfonts,nofootinbib,long]{revtex4}
\usepackage{epsfig,latexsym,bm}

\newcommand\GeV{\mbox{GeV}}

\newcommand\kpc{\mbox{kpc}}
\newcommand\Mpc{\mbox{Mpc}}
\newcommand\G{\mbox{G}}
\newcommand\A{\mathbf{A}}
\newcommand\B{\mathbf{B}}
\newcommand\E{\mathbf{E}}
\newcommand\x{\mathbf{x}}
\newcommand\y{\mathbf{y}}
\newcommand\z{\mathbf{z}}
\newcommand\kk{\mathbf{k}}

\newcommand\q{\mathbf{q}}
\newcommand\ee{{\boldsymbol \varepsilon}}
\newcommand\mPl{m_{\rm Pl}}
\newcommand\Ima{{\mbox{Im}}}
\newcommand\Rea{{\mbox{Re}}}

\begin{document}

\title{Helical Magnetic Fields from Inflation}

\author{Leonardo Campanelli$^{1,2}$}
\email{leonardo.campanelli@ba.infn.it}

\affiliation{$^1$Dipartimento di Fisica, Universit\`{a} di Bari, I-70126 Bari, Italy}
\affiliation{$^2$INFN - Sezione di Bari, I-70126 Bari, Italy}

\date{April, 2009}


\begin{abstract}
We analyze the generation of seed magnetic fields during de Sitter
inflation considering a non-invariant conformal term in the
electromagnetic Lagrangian of the form $-\frac14 I(\phi) F_{\mu
\nu} \widetilde{F}^{\mu \nu}$, where $I(\phi)$ is a pseudoscalar
function of a non-trivial background field $\phi$. In particular,
we consider a toy model, that could be realized owing to the
coupling between the photon and either a (tachyonic) massive
pseudoscalar field and a massless pseudoscalar field non-minimally
coupled to gravity, where $I$ follows a simple power-law behavior
$I(k,\eta) = g/(-k\eta)^{\beta}$ during inflation, while it is
negligibly small subsequently. Here, $g$ is a positive
dimensionless constant, $k$ the wavenumber, $\eta$ the conformal
time, and $\beta$ a real positive number. We find that only when
$\beta = 1$ and $0.1 \lesssim g \lesssim 2$ astrophysically
interesting fields can be produced as excitation of the vacuum,
and that they are maximally helical.
\end{abstract}


\maketitle
\newpage


\section{I. Introduction}

It is not excluded that large-scale, microgauss magnetic fields
detected in any type of galaxies have a primordial
origin~\cite{Widrow,Subramanian}. If so, they have probably been
generated during an inflationary epoch of the universe, since in
this case their correlation length can be as large as the galactic
one. (For possible generating mechanisms see, e.g.,
Ref.~\cite{Turner,Generation,Prokopec1,Campanelli1}.) An
overwhelming proof of ``primordial origin'' of galactic fields
could come from the observation of peculiar imprints they leave on
the Cosmic Microwave Background (CMB) anisotropies~\cite{CMB}.

Two considerations are in order. Firstly, to explain galactic
magnetism it suffices to generate ``seed'' magnetic fields prior to
galaxy formation of intensity generally much less than $1\mu \G$. In
fact, due to magnetohydrodynamic turbulence effects and differential
rotation of galaxy, extremely week fields can be exponentially
amplified. This mechanism, know as ``galactic
dynamo''~\cite{Dynamo,Subramanian}, successfully explains the main
characteristics of observed fields provided that their intensity and
correlation length are $B \gtrsim 10^{-33} \G$ and $\lambda \gtrsim
10 \kpc$.  If dynamo is inefficient, instead, a stronger field,
correlated on comoving scales of order $1 \Mpc$, is needed to
explain galactic magnetism. In this case the amplification of a
primordial seed field is just due to magnetic flux conservation
during protogalaxy collapses. Estimates based on ``spherical infall
model'' indicate that the time when protogalaxy collapses begins
corresponds to a redshift $z_{\rm pg}$ not greater than
$50$~\cite{Widrow}, although galactic disks are assembled at a much
later epoch, corresponding to redshifts of order $z \sim$ few.
Taking the ``less conservative'' value $z_{\rm pg} \simeq 50$, one
can shows that a comoving field $B \gtrsim 10^{-14} \G$ explain
magnetization of galaxies (see, e.g., Ref.~\cite{Campanelli1}).

Secondly, any viable mechanism of generation must repose on the
breaking of conformal invariance of standard electrodynamics,
otherwise the produced fields are vanishingly small~\cite{Turner}.

In this paper, we study the possibility to generate seed magnetic
fields during inflation in a non-invariant conformal theory of
electromagnetism described by the Lagrangian
\begin{equation}
\label{Lagrangian} {\cal L} = - \frac14 \, F_{\mu \nu} F^{\mu \nu}
- \frac14 \, I F_{\mu \nu} \widetilde{F}^{\mu \nu},
\end{equation}
where $F_{\mu \nu} = \partial_{\mu} A_{\nu} - \partial_{\nu}
A_{\mu}$ is the electromagnetic field strength tensor and
$\widetilde{F}_{\mu \nu} = (1/2\sqrt{-g}\,) \, \epsilon_{\mu \nu
\rho \sigma} F^{\rho \sigma}$
its dual, with $g$ the determinant of the metric tensor and
$\epsilon_{\mu \nu \rho \sigma}$ the Levi-Civita tensor.
The quantity $I$ is assumed to be a function of a non-trivial
background pseudoscalar field $\phi$ and its actual form will be
specified later.

In the seminal paper by Turner and Widrow~\cite{Turner}, it was
suggested that, during inflation, small fluctuating magnetic
fields could be amplified due to the coupling with the axion field
(in this case, the pseudoscalar $I$ is $I = g_{a\gamma} \phi$,
where $\phi$ is the axion field and $g_{a\gamma}$ is the
axion-photon coupling constant).
This possibility was exhaustively studied in the subsequent papers
by Garretson, Field and Carroll~\cite{Garretson}, and Field and
Carroll~\cite{Field}.
A different scenario was analyzed by Prokopec in
Ref.~\cite{Prokopec2} where the pseudoscalar $\phi$ was supposed
to drive (chaotic or extended) inflation.
As a matter of fact, in that analysis $I$ was taken to be a
uniform and slowly varying function of time during inflation and
zero subsequently.
In all cases, however, it was shown that no interesting cosmic
fields can be generated.

An interaction term of the form $I F_{\mu \nu} \widetilde{F}^{\mu
\nu}$ is responsible for the creation of magnetic fields described
by an ``unbalanced'' superposition of left- and right-handed
photons. In general, a time variation of $I$ produces a change of
$F_{\mu \nu} \widetilde{F}^{\mu \nu}$ which can be quantified
by~\cite{Campanelli2}
$\frac{1}{4V} \! \int_{t_1}^{t_2} \! d^4 x F_{\mu \nu}
\widetilde{F}^{\mu \nu} = H_V(t_2) - H_V(t_1)$. The quantity $H_V$
is the so-called magnetic helicity density (in the volume V) and
is proportional to the difference between the number of left- and
right-handed photons (see Section III). It is defined by
\begin{equation}
\label{H} H_V = \frac{1}{V} \! \int_V d^3x \A \cdot \B,
\end{equation}
where $\A$ and $\B$ are the vector potential and the magnetic
field, respectively.
In magnetohydrodynamics~\cite{Biskamp}, a non-vanishing magnetic
helicity indicates a non-trivial configuration of magnetic flux
tubes (which are twisted or linked).
In particle physics, magnetic helicity is known as the Abelian,
Euclidean Chern-Simons term. As shown by Jackiw and
Pi~\cite{Jackiw}, a helical magnetic field results from the
projection of a non-Abelian gauge field onto a fixed direction in
isospace. The magnetic helicity coincides then with the winding
number carried by the non-Abelian vacuum configuration.

It is worth noting that, since the magnetic helicity is odd under
discrete $P$ and $CP$ transformations, if detected on cosmic
scales it would indicate a macroscopic $P$ and $CP$ violation.
This peculiar characteristic of helical magnetic fields is very
attractive and, in fact, has induced many authors to devise models
of generation of primordial helical fields in the last
years~\cite{Turner,Garretson,Field,Prokopec2,Campanelli2,Jackiw,Cornwall,Giovannini1,
Forbes,Vachaspati1,Semikoz,Laine,Ahonen,Lee,Garcia-Bellido,Vachaspati2,Vachaspati3}.
Among the many mechanisms proposed, a very intriguing one is that
of Cornwall~\cite{Cornwall} and Vachaspati~\cite{Vachaspati1} (see
also Ref.~\cite{Vachaspati2,Vachaspati3,Garcia-Bellido}) according
to which a net helicity $H_V \sim - n_b/\alpha$ \cite{Vachaspati1}
is generated as a by-product of baryon-number-violating processes
taking place during electroweak baryogenesis, where $n_b$ is the
present cosmological baryon number density and $\alpha$ the fine
structure constant. Interestingly, as it has been pointed out by
Giovannini~\cite{Giovannini2}, the reverse it also true: decaying
non-trivial configuration of (hyper-)magnetic flux tubes can seed
the Baryon Asymmetry of the universe.

The plan of the paper is as follows. In Section II, we consider a
toy model which generalize the case analyzed by Prokopec in
Ref.~\cite{Prokopec2} and enables helical seed fields to be
produced as excitation of the vacuum.
In Section III, we compute the amount of magnetic helicity
associated to the generated fields.
In the Conclusions we summarize our results.


\section{II. Creation of Helical Magnetic Fields}

\subsection{a. General Considerations}

We assume that during inflation the universe is described by a de
Sitter metric
$ds^2 = a^2(d\eta^2 - d \x^2)$,
where $a(\eta)$ is the expansion parameter, $\eta = -1/(aH)$ is
the conformal time, and $H$ is the Hubble parameter. The conformal
time is related to the cosmic time $t$ through $d \eta = dt/a$. We
normalize the expansion parameter so that at the present time
$t_0$, $a(t_0) = 1$.

We work in the Coulomb gauge, $A_0 = \sum_{i=1}^3 \partial_i A_i
=0$, and we expand the electromagnetic field $A_{\mu} = (A_0, \A)$
as
\begin{eqnarray}
\label{Aexpansion0} && \A(\eta,\x) = \sum_{\alpha=1,2}
\A_\alpha(\eta,\x), \\
\label{Aexpansion} && \A_\alpha(\eta,\x) = \int \!\!
\frac{d^3k}{(2\pi)^3\sqrt{2k}} \: \ee_{\kk,\alpha} \,
a_{\kk,\alpha} A_{k,\alpha}(\eta) \, e^{i\kk \x} + \mbox{h.c.},
\end{eqnarray}
where $k = |\kk|$ and $\ee_{\kk,\alpha}$ (with $\alpha = 1,2$) are
the transverse polarization vectors satisfying the completeness
relation
$\sum_\alpha (\varepsilon_{\kk,\alpha})_i
(\varepsilon_{\kk,\alpha}^*)_j = \delta_{ij}- k_i k_j /k^2$.
The annihilation and creation operators $a_{\kk,\alpha}$ and
$a_{\kk,\alpha}^{\dag}$ satisfy the usual commutation relations
$[a_{\kk,\alpha}, a_{\kk',\alpha'}^{\dag}] = (2\pi)^3
\delta_{\alpha \alpha'} \delta(\kk-\kk')$,
$[a_{\kk,\alpha}, a_{\kk',\alpha'}] = [a_{\kk,\alpha}^{\dag},
a_{\kk',\alpha'}^{\dag}] = 0$,
with
$a_{\kk,\alpha} |0\rangle = 0$,
where $|0\rangle$ is the vacuum state normalized as $\langle
0|0\rangle = 1$.

Introducing the average magnetic and electric fields on a comoving
scale $\lambda$ as~\cite{Prokopec1}
\begin{eqnarray}
\label{Blambda0} && a^2 \B_{\lambda,\alpha}(\eta,\x) = \int \! d^3y
\, W_\lambda(|\x-\y|) \, \nabla \! \times \! \A_\alpha(\eta,\y),
\\
\label{Elambda0} && a^2 \E_{\lambda,\alpha}(\eta,\x) \! = - \int \!
d^3y \, W_\lambda(|\x-\y|) \, \dot{\A}_\alpha(\eta,\y),
\end{eqnarray}
where $W_\lambda(|\x|) = (2\pi \lambda^2)^{-3/2}
e^{-|\x|^2/(2\lambda^2)}$ is a gaussian window function and a dot
denotes differentiation with respect to the conformal time, we can
define their vacuum expectation values as
\begin{eqnarray}
\label{Blambda} && B_{\lambda,\alpha}^2(\eta) = \langle 0| \,
|\B_{\lambda,\alpha}(\eta,\x)|^2 |0 \rangle,
\\
\label{Elambda} && E_{\lambda,\alpha}^2(\eta) = \langle 0| \,
|\E_{\lambda,\alpha}(\eta,\x)|^2 |0 \rangle.
\end{eqnarray}
Taking into account Eqs.~(\ref{Aexpansion0})-(\ref{Elambda}), we
obtain
\begin{eqnarray}
\label{Blambda1} && B_{\lambda,\alpha}^2(\eta) = \int_{0}^{\infty}\!
\frac{dk}{k} \, W_\lambda^2(k) \, \mathcal{P}_{k,\alpha}(\eta),
\\
\label{Elambda1} && E_{\lambda,\alpha}^2(\eta) = \int_{0}^{\infty}\!
\frac{dk}{k} \, W_\lambda^2(k) \, \mathcal{Q}_{k,\alpha}(\eta),
\end{eqnarray}
where $W_\lambda(k) = e^{-\lambda^2 k^2/2}$ is the Fourier transform
of the window function and
\begin{eqnarray}
\label{PowerB} \mathcal{P}_{k,\alpha}(\eta) = \frac{k^4}{4\pi^2 a^4}
\, |A_{k,\alpha}(\eta)|^2,
\\
\label{PowerE} \mathcal{Q}_{k,\alpha}(\eta) = \frac{k^2}{4\pi^2 a^4}
\, |\dot{A}_{k,\alpha}(\eta)|^2,
\end{eqnarray}
are the magnetic and electric power spectra, respectively.

It is worth noting that in any meaningful model, the function
$\mathcal{P}_{k,\alpha}/k$ and $\mathcal{Q}_{k,\alpha}/k$ have to be
integrable in $k \rightarrow 0$ (no infrared divergence) in order to
have a finite value for the magnetic and electric fields
$B_{\lambda,\alpha}^2$ and $E_{\lambda,\alpha}^2$.

The equation of motion for $\A(\eta,\x)$ follows from
Lagrangian~(\ref{Lagrangian}):
\begin{equation}
\label{Motion0} \ddot{\A} - \nabla^2 \A + \dot{I} \, \nabla \times
\A - \nabla I \times \dot{\A} = 0,
\end{equation}
It is worth noting that the effect of the last term in the
left-hand-side of Eq.~(\ref{Motion0}) is just to cause a precession
of $\dot{\A}$ around $\nabla I$. Therefore, its presence does not
affect the intensity of $\dot{\A}$ and, in turn, that of ${\A}$ and
that of the related magnetic field. Retaining this term would add a
major level of complexity to our analysis without, nevertheless,
changing the final result regarding the average intensity of the
inflation-produced magnetic field. For this reason, we may write
\begin{equation}
\label{Motion0bis} \ddot{\A} - \nabla^2 \A + \dot{I} \, \nabla
\times \A \simeq 0,
\end{equation}
where the symbol $\simeq$ indicates equality in the sense just
discussed.

We are interested in the study of large-scale electromagnetic
fields, that is in modes whose physical wavelength is much greater
than the Hubble radius, $\lambda_{\rm phys} \gg H^{-1}$ or
equivalently $|k\eta| \ll 1$, where $\lambda_{\rm phys} = a \lambda$
and $\lambda = 1/k$ is the comoving wavelength.

If the quantity $I$ is peaked at some small wavelength we expect
that, at large scales, the term proportional to $I$ in
Eq.~(\ref{Motion0bis}) is negligible with respect to the second
term. In this case we recover the case of free Maxwell theory and no
amplification of magnetic modes occurs. Therefore, in the following,
we consider a simplified special case where $I$ is different from
zero just at large scales. To see how this assumption modifies
Eq.~(\ref{Motion0bis}), it is convenient to work in Fourier space.
Inserting Eqs.~(\ref{Aexpansion0})-(\ref{Aexpansion}) in
Eq.~(\ref{Motion0bis}), we get
\begin{equation}
\label{Motionk} \left( \frac{\ddot{A}_{k,\alpha}}{\sqrt{2k}} + k^2
\, \frac{A_{k,\alpha}}{\sqrt{2k}} \right) \! a_{\kk,\alpha} +
\sum_{\beta=1,2} \int \!\! \frac{d^3q}{(2\pi)^3} \, \dot{I}_{\kk -
\q} \Upsilon_{\kk,\alpha; \, \q,\beta} \,
\frac{A_{q,\beta}}{\sqrt{2q}} \: a_{\q,\beta} + \mbox{h.c.} \simeq
0,
\end{equation}
where $I_{\kk}$ is the Fourier transform of $I$ and
\begin{equation}
\label{Upsilon} \Upsilon_{\kk,\alpha; \, \q,\beta} = i \q \times
\ee_{\q,\beta} \cdot \ee_{\kk,\alpha}^* \, .
\end{equation}
Multiplying Eq.~(\ref{Motionk}) by $a_{\kk',\alpha'}^{\dag}$ first
on the right and then on the left and subtracting the resulting
expressions, and integrating in $\kk'$ and summing on $\alpha'$
afterwards, we obtain
\begin{equation}
\label{Motionkk} \frac{\ddot{A}_{k,\alpha}}{\sqrt{2k}} + k^2 \,
\frac{A_{k,\alpha}}{\sqrt{2k}} + \sum_{\beta=1,2} \int \!\!
\frac{d^3q}{(2\pi)^3} \, \dot{I}_{\kk - \q} \Upsilon_{\kk,\alpha;
\, \q,\beta} \, \frac{A_{q,\beta}}{\sqrt{2q}} \simeq 0.
\end{equation}
Since we are assuming that $I$ is different from zero just at large
scales ($k \rightarrow 0$), we may write
\begin{equation}
\label{Ik} I_{\kk} = (2\pi)^3 \delta(\kk) I_k,
\end{equation}
where $\delta(\kk)$ is a function which is ``extremely'' peaked at
small $k$. In order to simplify calculations, we assume that
$\delta(\kk)$ is indeed the Dirac delta function.
In this case, only modes with small wavenumbers are important when
considering the function $I_k$. For this reason, we may expand
$I_{\kk-\q}$ for small values of $\kk$ but, due to the presence of
the Dirac delta, we may expand for small values of $\q$ as well:
%
%
\begin{equation}
\label{Jexpansion} I_{\kk-\q} = (2\pi)^3 \delta(\kk-\q) [I_k - \q
\cdot \nabla_{\kk} I_k + \mathcal{O}(\q^{2})].
\end{equation}
Inserting the above expression into Eq.~(\ref{Motionkk}), we get
to the leading order
\begin{equation}
\label{Motion} \ddot{A}_{k,\alpha} + (k^2 \pm k\dot{I}_k)
A_{k,\alpha} \simeq 0,
\end{equation}
where we used
$\Upsilon_{\kk,\alpha; \, \kk,\beta} = \pm k \delta_{\alpha\beta}$
and, from now on, $\pm$ refer to $\alpha = 1,2$, respectively.

%
%
%
%

In the next three paragraphs, we will analyze the generation of
magnetic fields from the vacuum considering a simple
phenomenological form for the function $I_k$. In Section~II~e,
instead, we will give two examples of particle physics models in
which that form could naturally arise.

\subsection{b. A Simple Toy Model}

Let us suppose that $I_k$ follows a power-law behavior:
\begin{equation}
\label{I} I_k = \frac{g}{(-k\eta)^\beta} \, ,
\end{equation}
where $g$ is a dimensionless quantity (which for definiteness we
take to be positive) and we assume that $\beta > 0$.
[We note that the case analyzed by Prokopec in
Ref.~\cite{Prokopec2} corresponds to take $\beta \rightarrow 0$
and $g \beta = \mbox{const} \neq 0$ in Eq.~(\ref{I})].

For $|k\eta| \ll 1$ we can neglect the term proportional to $k^2$
in Eq.~(\ref{Motion}), so that its solution is
\footnote{The correct condition in order to neglect the term
proportional to $k^2$ in Eq.~(\ref{Motion}) is: $|k\eta| \ll
(\beta g)^{1/(\beta+1)}$. However, neglecting mathematical
quibbles such as the case where $g \rightarrow 0$ or $\beta
\rightarrow 0$, we assume that both $\beta$ and $g$ are quantities
of order unity, so the above relation reads $|k\eta| \ll 1$.}
\begin{equation}
\label{Solution} \beta \neq 1 \!: \, A_{k,\alpha}(\eta) =
|k\eta|^{1/2} \! \left[ c_1 H_{1/(1-\beta)}^{(1)}(z) + c_2
H_{1/(1-\beta)}^{(2)}(z) \right] \! ,
\end{equation}
where
\begin{equation}
\label{zeta} z = \frac{2\sqrt{\pm \beta g}}{1-\beta} \,
|k\eta|^{(1-\beta)/2},
\end{equation}
$c_1$ and $c_2$ are constants of integration, and $H_\nu^{(1)}(x)$
and $H_\nu^{(2)}(x)$ are the Hankel functions of first and second
kind, respectively. The case $\beta = 1$ will be analyzed
separately (see below).

Since $z \propto |k\eta|^{(1-\beta)/2}$, for $|k\eta| \ll 1$ we
have two different cases: if $0 < \beta < 1$ then $|z| \ll 1$,
while if $\beta > 1$ then $|z| \gg 1$. Consequently, in
Eq.~(\ref{Solution}) we can use the asymptotic expansion of the
Hankel functions for small and large arguments, respectively:
\begin{equation}
\label{Hankel} H_\nu^{(1,2)}(x) \simeq
    \left\{ \begin{array}{lll}
        \mp \frac{i 2^\nu \!\csc(\pi \nu)}{\Gamma(1-\nu)} \: x^{-\nu}
        +
        \frac{2^{-\nu} [1 \pm i \! \cot(\pi \nu)]}{\Gamma(1+\nu)} \:
        x^{\nu},                              & |x| \ll 1, \nu \neq 0,
        \\ \\
        \pm 2i \pi^{-1} \ln x,                & |x| \ll 1, \nu = 0,

        \\ \\
        \sqrt{\frac{2}{\pi x}} \: e^{\pm
        i(x-\frac{\pi}{4}-\frac{\pi\nu}{2})}, & |x| \gg 1.
            \end{array}
    \right.
\end{equation}
%
%
If $0 < \beta < 1$, using the first equation of (\ref{Hankel}), we
find $A_{k,\alpha}(\eta) \simeq c'_1 + c'_2 |k\eta|$, where $c'_1$
and $c'_2$ are constants. Since for modes well inside the horizon
($|k\eta| \rightarrow \infty$) we have the plane-wave solution
$A_{k,\alpha}(\eta) = e^{ik\eta}$ (the normalization corresponds
to the standard Bunch-Davies vacuum~\cite{Birrell}), and for modes
well outside the horizon we have the above solution, we can fix
the values of $c'_1$ and $c'_2$ by matching the two solutions and
their first derivatives at the horizon crossing, $|k\eta| = 1$. We
find $c'_1 = (1+i)e^{-i}$ and $c'_2 = -ie^{-i}$, so as a final
result we can write the expression for the electromagnetic field
during inflation and at large scales:
\footnote{It is worth noting that the matching procedure we have
adopted can give us only approximate results. Nevertheless, we
believe that the simplified analysis performed in the case $0 <
\beta < 1$ catches the main characteristics of the process of
creation of magnetic field during inflation.}
\begin{equation}
\label{Ainflation1} 0 < \beta < 1\!: \, A_{k,\alpha}(\eta) \simeq
e^{-i} \, [\, 1 + i(1-|k\eta|) \,].
\end{equation}
If $\beta > 1$, using the third equation of (\ref{Hankel}), we
find
$A_{k,\alpha}(\eta) \simeq |k\eta|^{(1+\beta)/2} (c''_1 e^{iz} +
c''_2 e^{-iz})$,
where $c''_1$ and $c''_2$ are constants. We see that the positive
helical modes [corresponding to take the plus sign in
Eq.~(\ref{zeta})] are $A_{k,+}(\eta) \propto
|k\eta|^{(1+\beta)/2}$ (we neglected inessential oscillating
factors), while the negative ones are $A_{k,-}(\eta) \propto
|k\eta|^{(1+\beta)/2} e^{|z|}$. Therefore, the spectrum of
positive helicity states is vanishingly small, while that of
negative helicity states presents an infrared divergence
(associated to the exponential factor). Therefore, the case $\beta
> 1$ is meaningless and then will be neglected in the following.

In the case $\beta = 1$, Eq.~(\ref{Motion}) can be solved exactly
(for all $|k\eta|$):
\begin{equation}
\label{Ainflation} \beta = 1\!: \, A_{k,\alpha}(\eta) =
\sqrt{\frac{\pi}{2}} \, e^{-i\pi(1+2\nu)/4} \, |k\eta|^{1/2} \,
H_\nu^{(2)}(|k\eta|),
\end{equation}
where
\begin{equation}
\label{nu} \nu = \sqrt{\frac14 \mp g}
\end{equation}
and we used the normalization corresponding to the Bunch-Davies
vacuum.

For large scales, we can replace the Hankel function with its
small-argument expansion in Eq.~(\ref{Ainflation}).

\subsection{c. Electromagnetic Backreaction on Inflation}

Before proceeding further, we want to analyze the problem of
backreaction of the inflation-produced electromagnetic field on the
dynamics of inflation.
\footnote{After inflation, since the conductivity of the cosmic
plasma becomes very high and, consequently, the electric field is
washed out~\cite{Turner}, the electromagnetic energy is simply given
by the magnetic one. Moreover, the magnetic field evolves
adiabatically from the end of inflation until today, and its energy
density is always subdominant with respect to the energy density of
the universe (strictly speaking, this is true for cosmic magnetic
fields whose intensity is consistent with astrophysical
observations, that is $B \lesssim 10^{-9} \G$). This means that we
can safely neglect the backreaction of the electromagnetic field on
the standard evolution of the universe from inflation until today.}
Since in the following we will assume that the electromagnetic field
does not appreciably perturb the standard evolution of the universe,
we must verify that the electromagnetic energy density is smaller
than the energy density associated to inflation which, during de
Sitter inflation, is a constant: $\rho_{\rm infl} = T_1^4$. Here,
$T_1$ is the so-called reheating temperature, that is the
temperature of the cosmic plasma at the beginning of the radiation
era (here and in the following we assume that the reheating phase,
during which the energy of the inflaton is converted into ordinary
matter is ``instantaneous'' so that, after inflation, the universe
enters directly the radiation era).

Starting from the definition of the energy-momentum tensor,
\begin{equation}
\label{Tmunu1} T_{\mu \nu} = \frac{2}{\sqrt{-g}} \, \frac{\delta
S}{\delta g^{\mu \nu}} \, ,
\end{equation}
where $S = \int \! d^4 x \sqrt{-g} \, \mathcal{L}$ is the action,
and taking into account Lagrangian~(\ref{Lagrangian}), we find
\begin{equation}
\label{Tmunu2} T_{\mu \nu} = \frac14 g_{\mu \nu} F_{\alpha \beta}
F^{\alpha \beta} - F_{\mu}^{\;\; \alpha} F_{\nu \alpha},
\end{equation}
that is the standard Mawxell energy-momentum tensor for the
electromagnetic field. Here, we have assumed, for the sake of
simplicity, that the quantity $I$ does not explicitly depend on the
metric tensor $g_{\mu \nu}$ (two particle physics models in which
$I$ displays this property are discussed in Section II e). The
electromagnetic energy density, $\rho = T_0^0$, is then given by
\begin{equation}
\label{Energy1} \rho = \frac12 (\E^2 + \B^2),
\end{equation}
where $a^2 \B = \nabla \times \A$ and $a^2 \E = -\dot{\A}$.
Expanding the electromagnetic field as in
Eqs.~(\ref{Aexpansion0})-(\ref{Aexpansion}), replacing $\B_{\alpha}$
and $\E_{\alpha}$ with their average values $\B_{\lambda,\alpha}$
and $\E_{\lambda,\alpha}$, and then taking the vacuum expectation
value of expression (\ref{Energy1}), we obtain the vacuum
expectation value of the average electromagnetic energy density on a
comoving scale $\lambda$:
\begin{equation}
\label{Energy2} \rho_{\lambda} = \frac12 \sum_{\alpha=1,2} \left(
B_{\lambda,\alpha}^2 + E_{\lambda,\alpha}^2 \right).
\end{equation}
In order to have a self-consistent model for the production of a
cosmic magnetic field, we must verify that, on the scale of interest
$\lambda$, the electromagnetic energy produced during inflation is
less than the total energy of the universe:
\begin{equation}
\label{Energy3} \rho_\lambda < \rho_{\rm infl}.
\end{equation}
Taking into account Eqs.~(\ref{Blambda})-(\ref{PowerE}) and
Eqs.~(\ref{Ainflation1}), (\ref{Ainflation}), and (\ref{Energy2}),
we easily find, at large scales:
\begin{equation}
\label{rho1} 0 < \beta < 1\!: \, \rho_{\lambda}(\eta) \simeq \frac{3
H^4}{8\pi^2} \left( \frac{\eta}{\lambda} \right)^{\!4}
\end{equation}
and
\begin{equation}
\beta = 1\!: \, \rho_{\lambda}(\eta) \simeq
          \frac{H^4}{32 \pi^{3}} \, 4^{\nu_{+}} \!\left( \frac12 -\nu_{+} \right)^{\!\!2}
          [\Gamma(\nu_{+})]^2 \: \Gamma \! \left( \frac32-\nu_{+} \! \right) \!
          \left( \frac{|\eta|}{\lambda} \right)^{\!\!3-2\nu_{+}} \!\! ,
\label{rho2}
\end{equation}
where $\nu_{+} = \sqrt{\frac14 + g}$. It is important to observe
that, in order to avoid an infrared divergence in the magnetic power
spectrum, we must impose that $\nu < 3/2$.

Taking into account that $T_1^4 \simeq H^2 \mPl^2$~\cite{Turner},
where $\mPl \simeq 1.22 \times 10^{19} \GeV$ is the Planck mass, we
have approximatively
\begin{equation}
\label{r1} 0 < \beta < 1\!: \, \frac{\rho_{\lambda}}{\rho_{\rm infl}} \sim
          10^{-10} \left( \frac{T_1}{10^{-2} \mPl} \right)^{\!\!4}
          \left( \frac{|\eta|}{\lambda} \right)^{\!\!4}
\end{equation}
and
\begin{equation}
\label{r2} \beta = 1\!: \, \frac{\rho_{\lambda}}{\rho_{\rm infl}} \sim
          10^{-10} \frac{f(\nu_{+})}{3-2\nu_{+}} \left( \frac{T_1}{10^{-2} \mPl} \right)^{\!\!4}
          \left( \frac{|\eta|}{\lambda} \right)^{\!\!3-2\nu_{+}} \!\! ,
\end{equation}
where $f(x)$ is an increasing function of $x$, such that $f(1/2) =
0$ and $f(3/2) \sim 1$.  Since analyses of CMB radiation and Big
Bang Nucleosynthesis constrain the reheating temperature in the
range $1\GeV \lesssim T_1 \lesssim 10^{-2}\mPl$~\cite{Turner}, and
since we are considering the case of large scales, $|\eta|/\lambda
\ll 1$, we see form Eqs.~(\ref{r1}) and (\ref{r2}) that we can
safely neglect the backreaction of the electromagnetic field on the
dynamics of inflation.

\subsection{d. The Actual, Inflation-produced, Magnetic Field}

After inflation, the universe enters the radiation era (assuming
instantaneous reheating). We restrict our analysis to the case in
which, during this era, the interaction term in
Lagrangian~(\ref{Lagrangian}) is (for some reason) negligible, so
that the general expression for the electromagnetic field is
\begin{equation}
\label{Aradiation} A_{k,\alpha}^{\rm rad}(\eta) =
\alpha_{k,\alpha} e^{ik\eta} + \beta_{k,\alpha} e^{-ik\eta}.
\end{equation}
Here, $\alpha_{k,\alpha}$ and $\beta_{k,\alpha}$ are the so-called
Bogoliubov coefficients~\cite{Birrell}, determining the spectral
number distribution of particles produced from the vacuum. By
matching expressions (\ref{Ainflation1}) and (\ref{Aradiation})
and their first derivatives at the time of the end of inflation,
$\eta = \eta_1$, we find the spectrum of the electromagnetic field
generated from the vacuum at large scales:
\begin{equation}
\label{Ageneratedtris} 0 < \beta < 1 \!: \, |A_{k,\alpha}^{\rm
vac}(\eta_1)|^2 = |\beta_{k,\alpha}|^2 \simeq \frac14 \, .
\end{equation}
In the same way, we find:
\begin{equation}
\beta = 1\!: \, |A_{k,\alpha}^{\rm vac}(\eta_1)|^2 \simeq
    \left\{ \begin{array}{lll}
          \frac{[2^{\nu} \! \left( \frac12-\nu \right)
          \Gamma(\nu)]^2}{8\pi |k\eta_1|^{1+2\nu}},                             & \nu > 0,
        \\ \\
          \frac{(\ln|k\eta_1|)^2}{8\pi |k\eta_1|},                              & \nu = 0,
        \\ \\
          \frac{a_\nu + b_\nu \!
          \cos[2\tilde{\nu}\ln(|k\eta_1|/2)]}{4\pi |k\eta_1|} +
          \frac{c_\nu \! \sin[2\tilde{\nu}\ln(|k\eta_1|/2)]}{4\pi |k\eta_1|},   & \nu = i\tilde{\nu}.
    \end{array}
    \right.
\label{Agenerated}
\end{equation}
%
%
Here, $\tilde{\nu} > 0$ is the imaginary part of $\nu$ and
\begin{eqnarray}
&& a_\nu = \frac{\pi}{\tilde{\nu}} \, (1+4\tilde{\nu}^2)
\coth(\pi\tilde{\nu}), \nonumber \\
&& b_\nu =(1-4\tilde{\nu}^2) \Rea[\Gamma(\nu)^2] + 4\tilde{\nu}
\Ima[\Gamma(\nu)^2], \\
&& c_\nu =(1-4\tilde{\nu}^2) \Ima[\Gamma(\nu)^2] - 4\tilde{\nu}
\Rea[\Gamma(\nu)^2], \nonumber
\end{eqnarray}
%
%
%
%
where $\Rea[x]$ and $\Ima[x]$ are the real and imaginary part of
$x$.

We observe that when $g=0$ ($\nu = 1/2$) conformal invariance is
recovered and then, as it should be, we find $|A_{k,\alpha}^{\rm
vac}(\eta_1)|^2 = 0$.

From the end of inflation until today, due to the high
conductivity of the cosmic plasma, the magnetic field evolves
adiabatically~\cite{Turner}, $a^2 B_{\lambda,\alpha} =
\mbox{const}$, so that $B_{\lambda,\alpha}^{\rm today} = a_1^2
B_{\lambda,\alpha}(\eta_1)$. Inserting Eqs.~(\ref{Ageneratedtris})
and (\ref{Agenerated}) in Eq.~(\ref{Blambda1}), we obtain
respectively
\begin{equation}
\label{B0tris} 0 < \beta < 1\!: \, B_{\lambda,\alpha}^{\rm today}
\simeq \frac{1}{4\sqrt{2}\pi \lambda^{2}} \, ,
\end{equation}
and
\begin{equation}
\label{B0} \!\!\! \beta = 1\!: \, B_{\lambda,\alpha}^{\rm today}
\simeq
    \left\{ \begin{array}{lll}
          \!\!\! \frac{2^\nu |\frac12-\nu| \,
          \Gamma(\nu) \, [\Gamma(\frac32-\nu)]^{1/2}}{(4\pi)^{3/2} \lambda^{2}}
          \! \left(\frac{\lambda}{|\eta_1|} \right)^{\!\nu + 1/2}                  & \nu > 0,
        \\ \\
          \!\!\! \frac{\ln(\lambda/|\eta_1|)}{2(4\pi)^{5/4} \lambda^{2}}
          \! \left(\frac{\lambda}{|\eta_1|} \right)^{\!\!1/2}                      & \nu = 0,
        \\ \\
          \!\!\! \frac{(a_\nu + b_\nu/\sqrt{2} +
          c_\nu/\sqrt{2})^{1/2}}{\sqrt{2}(4\pi)^{5/4} \lambda^{2}}
          \! \left(\frac{\lambda}{|\eta_1|} \right)^{\!\!1/2}                      & \nu = i\tilde{\nu},
    \end{array}
    \right.
\end{equation}
%
%
where, for simplicity, we replaced in the third equation of
(\ref{Agenerated}) the cosine and sine functions with their
root-mean-square values $1/\sqrt{2}$.
%
%

It is useful to observe that
\begin{eqnarray}
\label{lambdaeta} && \frac{\lambda}{|\eta_1|} \simeq 1.22 \times
10^{23} \, \frac{T_1}{\mPl} \, \lambda_{\rm 10 kpc}, \\
\label{lambda2} && \lambda^{-2} \simeq 0.60 \times 10^{-53}
\lambda_{\rm 10 kpc}^{-2} \G,
\end{eqnarray}
where $\lambda_{\rm 10 kpc} = \lambda/(10 \kpc)$. In
Eq.~(\ref{lambdaeta}), we used the fact that during radiation and
matter dominated eras the expansion parameter evolves as $a \propto
g_{*S}^{-1/3} T^{-1}$, where $T$ is the temperature and $g_{*S}(T)$
the number of effectively massless degrees of freedom referring to
the entropy density of the universe~\cite{Kolb}.
\footnote{In this paper, we use the values~\cite{Kolb}: $T_0
\simeq 2.35 \times 10^{-13} \GeV$, $g_{*S}(T_0) \simeq 3.91$, and
$g_{*S}(T_1) = 106.75$ (referring to the number of effectively
massless degrees of freedom of Standard Model). It is useful to
know that $1\G \simeq 6.9 \times 10^{-20} \GeV^2$ and $1 \Mpc
\simeq 1.56 \times 10^{38} \GeV^{-1}$.}

Taking into account Eqs.~(\ref{B0tris}) and
(\ref{lambdaeta})-(\ref{lambda2}), we see that magnetic fields
produced in the cases $0 < \beta < 1$ are well below the minimum
seed field required for dynamo amplification, $B \simeq 10^{-33}
\G$.

In Fig.~1, we show the actual magnetic field in the case $\beta =
1$ as a function of the parameter $g$ at the comoving scales
$\lambda = 10\kpc$ (thin lines) and $\lambda = 1\Mpc$ (thick
lines), for $T_1 = 10^{-2} \mPl$. Dashed and continuous lines
refer to the cases $\alpha = 1$ (positive helicity states) and
$\alpha = 2$ (negative helicity states), respectively.


\begin{figure}[t]
\begin{center}
\includegraphics[clip,width=0.7\textwidth]{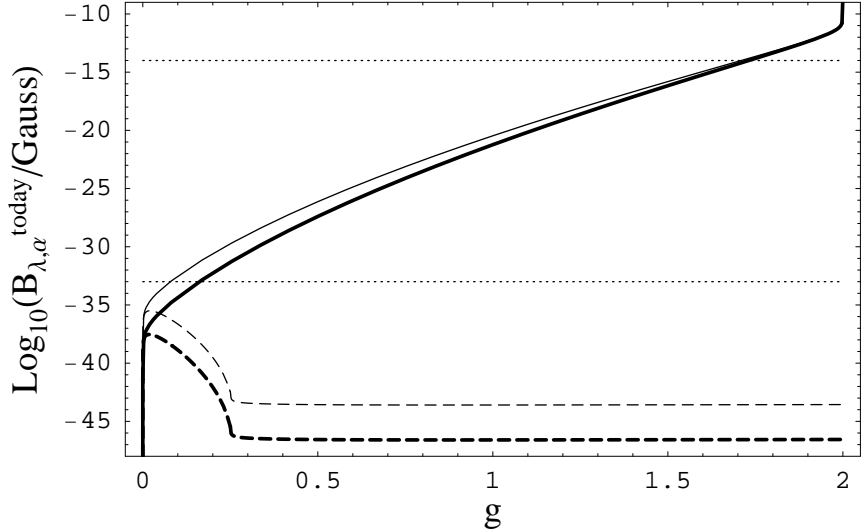}
\caption{Actual magnetic field in the case $\beta = 1$ as a
function of $g$ at the comoving scales $\lambda = 10\kpc$ (thin
lines) and $\lambda = 1\Mpc$ (thick lines), for a reheating
temperature $T_1 = 10^{-2} \mPl$. Dashed and continuous lines
refer to the cases $\alpha = 1$ (positive helicity states) and
$\alpha = 2$ (negative helicity states), respectively. The
horizontal dotted lines refer to the minimum seed fields required
for dynamo amplification, $B \simeq 10^{-33} \G$, and to directly
explain galactic magnetism, $B \simeq 10^{-14} \G$.}
\end{center}
\end{figure}


As it is clear from the figure, astrophysically interesting fields
can be produced as excitation of the vacuum. They have a definite
helicity
\footnote{Since we are considering $g \geq 0$, only negative
helicity states are astrophysically relevant. Had we taken $g \leq
0$, the non-vanishing modes would have been the positive ones.}
and intensity strongly depending on the value of $g$. (The case $g
\geq 2$, or equivalently $\nu > 3/2$, corresponds to a
non-physical infrared-divergent magnetic power spectrum).
Numerically we find that for $g \gtrsim 0.1$ the produced field on
scales of $10\kpc$ is stronger than the minimum seed field
required for a successful galactic dynamo amplification, while if
$1.7 \lesssim g < 2$ its intensity on scales of $1\Mpc$ is high
enough to directly explain galactic magnetism ($B \gtrsim 10^{-14}
\G$).
In the range of interest,
$0.1 \lesssim g \lesssim 2$, we have approximately
\begin{equation}
\label{B0-estimate} B_{\lambda,-}^{\rm today} \sim \:
10^{23\nu-47} \left( \frac{T_1}{10^{-2}\mPl} \right)^{\!\!\nu+1/2}
\lambda_{\rm 10 kpc}^{\nu-3/2} \; \G,
\end{equation}
where we used the first equation of (\ref{B0}) with $\alpha = 2$,
and Eqs.~(\ref{lambdaeta})-(\ref{lambda2}).

\subsection{e. Examples of Particle Physics Models}

We want now give two particle physics models in which the function
$I_{\kk}$ takes on the form~(\ref{Ik}) with $I_k$ given by
Eq.~(\ref{I}).

Let us start by considering a model in which $I$ is given by $I =
\phi/M$, where $M$ is a mass scale and $\phi$ a pseudoscalar
(background) field whose Lagrangian density is
\begin{equation}
\label{Lphi} \mathcal{L}_{\phi} = \frac12 \partial_{\mu} \phi \,
\partial^{\mu} \phi + V(\phi),
\end{equation}
with $V(\phi)$ a potential term that will be specified later.

In the spirit of mean field theory, we may approximate the
function $\phi(\x,\eta)$ by its root-mean-square value
$\phi_\lambda(\eta)$ on the scale $\lambda$:
\begin{equation}
\phi_\lambda^2(\eta) = \langle 0| \, |\int \!\! d^3 y
W_\lambda(|\x-\y|) \phi(\y,\eta)|^2 \, |0\rangle,
\end{equation}
where $W_\lambda(|{\mathbf{x}}|)$ is a suitable window function.
Introducing the $\phi$-power spectrum,
$\mathcal{P}_{\phi}(k,\eta)$, through the relation
\begin{equation}
\langle 0| \phi_{\mathbf{k}}(\eta) \phi_{\mathbf{q}}^*(\eta)
|0\rangle = (2\pi)^3 \delta^3(\mathbf{k}-\mathbf{q})
\frac{2\pi^2}{k^3} \mathcal{P}_{\phi}(k,\eta),
\end{equation}
where $\phi_{\mathbf{k}}$ is the Fourier transform of $\phi(\x)$,
it is straightforward to obtain
\begin{equation}
\phi_\lambda^2(\eta) = \int_0^{\infty} \!\! \frac{dk}{k} \,
|W_\lambda(k)|^2 \, \mathcal{P}_{\phi}(k,\eta),
\end{equation}
where $W_\lambda(k)$ is the Fourier transform of the window
function. For the sake of simplicity, let us assume that
$|W_\lambda(k)|^2$ picks out just modes with wavenumber $k =
1/\lambda$. In other words, we take $|W_\lambda(k)|^2 =
\delta(k\lambda -1)$, where $\delta(x)$ is the Dirac delta
function. Then, we may write:
$\phi_\lambda^2(\eta) = \mathcal{P}_{\phi}(1/\lambda,\eta)$.
Consequently, the Fourier transform of $I({\mathbf{x}},\eta)$ is
simply given by:
\begin{equation}
\label{x} I_{\kk}(\eta) = (2\pi)^3 \delta^3(\kk) \,
\frac{[\mathcal{P}_{\phi}(1/\lambda,\eta)]^{1/2}}{M} \, .
\end{equation}
To proceed further, we assume that the $\phi$-power spectrum is
peaked at large scales, $\lambda \rightarrow \infty$. Taking into
account the presence of the Dirac function in Eq.~(\ref{x}), we
can roughly write
$I_{\kk}(\eta) \simeq (2\pi)^3\delta^3(\kk)
[\mathcal{P}_{\phi}(k,\eta)]^{1/2}\!/M$,
since both $k$ and $1/\lambda$ go to zero. Therefore, the function
$I_k$ introduced in Eq.~(\ref{Ik}) takes on the form
\begin{equation}
\label{x1} I_k(\eta) \simeq
\frac{[\mathcal{P}_{\phi}(k,\eta)]^{1/2}}{M} \, .
\end{equation}
To be more concrete, let us now consider the case in which $\phi$
is either a massive pseudoscalar field minimally coupled to
gravity and a massless pseudoscalar field non-minimally coupled to
gravity. They are described by Lagrangian~(\ref{Lphi}) with
%
\begin{equation}
V(\phi) = \left\{
               \begin{array}{lll}
                m^2 \phi^2,
                \\
               \xi R \phi^2,
    \end{array}
    \right.
\end{equation}
respectively. Here, $m^2$ is the squared mass of the pseudoscalar
field, $\xi$ a real parameter, and $R$ the Ricci scalar. It is
well-known that in those cases, the $\phi$-power spectrum in de
Sitter background is given by (see, e.g., Ref.~\cite{Riotto}):
\begin{equation}
\mathcal{P}_{\phi}(k,\eta) = \left( \frac{H}{2\pi} \right)^{\!\!2}
(-k\eta)^{3-2\nu_\phi},
\end{equation}
where 
%
\begin{equation}
\nu_\phi = \left\{
                \begin{array}{lll}
                 \sqrt{\frac94 -\frac{m^2}{H^2}} \: ,
                 \\
                 \sqrt{\frac94 - 12 \, \xi} \: ,
           \end{array}
    \right.
\end{equation}
in the two cases, respectively. Comparing Eq.~(\ref{x1}) with
Eq.~(\ref{I}), we finally get
\begin{eqnarray}
&& g = \frac{H}{2\pi M} \, ,
\\
&& \beta = \nu_\phi - \frac32 \, .
\end{eqnarray}
It is clear that only negative values of $m^2$ and $\xi$ give
positive values of $\beta$. Moreover, the astrophysically
interesting case discussed in the previous Section, $\beta = 1$
and $0.1 \lesssim g \lesssim 2$, corresponds to have $m^2 = -4H^2$
and $\xi = -1/3$, and $0.1 H \lesssim M \lesssim 1.6 H$.


\section{III. Creation of Magnetic Helicity}

Let us now consider the creation of magnetic helicity in our
model. The magnetic helicity in a volume $V = \lambda^3$ can be
conveniently defined as
\begin{equation}
\label{H1} H_\lambda(\eta) = \langle 0| \!\int \!\! d^3y \!\! \int
\!\! d^3z W_\lambda(|\x-\y|) W_\lambda(|\x-\z|) \A(\eta,\y) \cdot
\B(\eta,\z) |0 \rangle,
\end{equation}
where
$W_\lambda(|\x|)$ is a gaussian window function.
Taking into account Eqs.~(\ref{Aexpansion0})-(\ref{Aexpansion}),
we obtain
%
%
%
\begin{equation}
\label{H2} H_\lambda(\eta) = \int_{0}^{\infty} \! \frac{dk}{k} \,
W_\lambda^2(k) \, \mathcal{H}_k(\eta),
\end{equation}
where
\begin{equation}
\label{PowerH} \mathcal{H}_k(\eta) = \frac{k^3}{4\pi^2 a^2} \left[
\, |A_{k,+}(\eta)|^2 - |A_{k,-}(\eta)|^2 \, \right]
\end{equation}
is the magnetic helicity power spectrum (which is well defined
only when $\mathcal{H}_k/k$ is integrable in $k \rightarrow 0$).

A magnetic field is said to be ``maximally helical'' at the time
$\bar{\eta}$ if either $A_{k,+}(\bar{\eta})$ or
$A_{k,-}(\bar{\eta})$ is zero.

Looking at Eqs.~(\ref{Ageneratedtris}) and (\ref{PowerH}), we
immediately get that in the case $0 < \beta < 1$ the magnetic
helicity is zero. In the case $\beta=1$, instead, taking into
account Eq.~(\ref{Agenerated}), we obtain
\begin{equation}
H_\lambda^{\rm today} \simeq
    \left\{ \begin{array}{ll}
        \frac{\gamma_{+}}{\sqrt{\pi}} \,
        (B_{\lambda,+}^{\rm today})^2 \lambda -
        \frac{\gamma_{-}}{\sqrt{\pi}} \,
        (B_{\lambda,-}^{\rm today})^2 \lambda,      & 0 \leq g \leq \frac14,
        \\ \\
        \frac{2}{\sqrt{\pi}} \,
        (B_{\lambda,+}^{\rm today})^2 \lambda -
        \frac{\gamma_{-}}{\sqrt{\pi}} \,
        (B_{\lambda,-}^{\rm today})^2 \lambda,      & \frac14 \leq g < \frac34,
    \end{array}
    \right.
\label{Htoday}
\end{equation}
%
%
%
%
%
%
where
$\gamma_{\pm} = \mathrm{B}(1/2 \, , 1 - \nu)$
%
%
and $\mathrm{B}(x,y)$ is the Euler beta function. (The case $g
\geq 3/4$ corresponds to an infrared-divergent magnetic helicity
power spectrum.)

It is worth noting that, al least in the case of astrophysical
interest, the generated magnetic fields possess a definite
helicity (in particular they are left-handed since $|A_{k,+}| \ll
|A_{k,-}|\,$). Hence, they are (almost) maximally helical, and the
amount of created magnetic helicity is roughly given by
\begin{equation}
\label{Htoday1} H_\lambda^{\rm today} \sim - (B_{\lambda,-}^{\rm
today})^2 \lambda .
\end{equation}
A helical magnetic field leaves peculiar imprints on the CMB
radiation. Unfortunately, the maximal helicity producible in our
mechanism is much smaller than that detectable in near future CMB
experiments which is, in the most optimistic case, of order of
$(10^{-9} \G)^2 \Mpc$ on $\Mpc$ scales~\cite{Caprini}.
%
%
In fact, at those scales we get from Eq.~(\ref{Htoday1}) and
Fig.~1 the value $|H_\lambda^{\rm today}|_{\rm max} \sim (10^{-24}
\G)^2 \Mpc$.

Interesting enough, it has been showed by Kahniashvili and
Vachaspati~\cite{Kahniashvili} that, in principle, the study of
propagation properties of charged cosmic rays through a magnetic
field could give information about the helicity of the field
itself.


\section{IV. Conclusions}

Up to today, the hypothesis that {\it all} galaxies are pervaded
by microgauss magnetic fields has not been ruled out yet.
Due to their large correlation length, we can entertain the idea
that they are remnant of an inflationary epoch of the universe.

In this paper, we have indeed analyzed the production of seed
magnetic fields during de Sitter inflation. We have considered a
photon interaction term in the electromagnetic Lagrangian (which
breaks conformal invariance of electrodynamics) of the form
$-\frac14 I F_{\mu \nu} \widetilde{F}^{\mu \nu}$, where $I$ is a
pseudoscalar function.
We have considered a simplified special case where $I$ is peaked
at large scales ($k = 1/\lambda \rightarrow 0$) during inflation
and then is vanishingly small afterwards. In particular, we have
studied the case where $I$ is parameterized by the power-law
behavior $I(k,\eta) = g/(-k\eta)^\beta$, with $g$ a positive
dimensionless constant, $\beta$ a real positive number, and $\eta$
the conformal time. Also, we have shown that this particular form
of $I$ could be realized owing to the coupling between the photon
and either a (tachyonic) massive pseudoscalar field and a massless
pseudoscalar field non-minimally coupled to gravity.

We have found that only when $\beta = 1$ astrophysically
interesting fields can be produced as excitation of the vacuum.
They are maximally helical and have an intensity strongly
depending on the value of $g$ (which, in order to avoid an
infrared divergence in the magnetic power spectrum, is constrained
in the range $0 \leq g < 2$).
In particular, for $g \gtrsim 0.1$ the produced fields on scales
of $10\kpc$ are stronger than the minimum seed field required for
a successful galactic dynamo amplification ($B \gtrsim 10^{-33}
\G$), while if $1.7 \lesssim g < 2$ their intensity on scales of
$1\Mpc$ is high enough to directly explain galactic magnetism ($B
\gtrsim 10^{-14} \G$).

\newpage


\end{document}